\documentclass[letterpaper, 10 pt, conference]{ieeeconf}  

\IEEEoverridecommandlockouts                              

\usepackage{graphicx,algorithm,algorithmic,amsthm,amsmath,amsfonts,verbatim,mathtools,enumerate,bm,hyperref,xcolor,mathabx,enumerate,arydshln}
\makeatletter
\renewcommand*\env@matrix[1][*\c@MaxMatrixCols c]{%
  \hskip -\arraycolsep
  \let\@ifnextchar\new@ifnextchar
  \array{#1}}
\makeatother

\allowdisplaybreaks

\DeclareMathOperator{\im}{im}

\newcommand{\norm}[1]{\left\lVert#1\right\rVert}
\newcommand{\mnorm}[1]{{\left\vert\kern-0.25ex\left\vert\kern-0.25ex\left\vert #1 
    \right\vert\kern-0.25ex\right\vert\kern-0.25ex\right\vert}}

\newtheorem{definition}{Definition} 
\newtheorem{theorem}{Theorem}

\newtheorem{corollary}{Corollary}
\newtheorem{remark}{Remark}

\newcommand{\ie}{{\it i.e.}}

\usepackage{todonotes}

\title{\LARGE \bf On Controllability and Persistency of Excitation in Data-Driven Control: Extensions of Willems' Fundamental Lemma
}

\author{Yue Yu, Shahriar Talebi, Henk J. van Waarde, Ufuk Topcu, Mehran Mesbahi, and Beh\c{c}et~A\c{c}\i kme\c{s}e
\thanks{Yue Yu and Ufuk Topcu are with the Department of Aerospace Engineering, University of Texas at Austin, Austin, TX 78712 USA. Shahriar Talebi, Mehran Mesbahi and Beh\c{c}et~A\c{c}\i kme\c{s}e are with the Department of Aeronautics and Astronautics, University of Washington, Seattle, WA 98195 USA. Henk J. van Waarde is with the Control Group, Department of Engineering, University of Cambridge, Trumpington Street, Cambridge CB2 1PZ, UK.  (emails:yueyu@utexas.edu, shahriar@uw.edu, hv280@cam.ac.uk, utopcu@utexas.edu, mesbahi@uw.edu, behcet@uw.edu)}
}

\begin{document}

\maketitle
\thispagestyle{empty}
\pagestyle{empty}

\begin{abstract}
Willems' fundamental lemma asserts that all trajectories of a linear time-invariant system can be obtained from a finite number of measured ones, assuming that controllability and a persistency of excitation condition hold. We show that these two conditions can be relaxed. First, we prove that the controllability condition can be replaced by a condition on the controllable subspace, unobservable subspace, and a certain subspace associated with the measured trajectories. Second, we prove that the persistency of excitation requirement can be relaxed if the degree of a certain minimal polynomial is tightly bounded. Our results show that data-driven predictive control using online data is equivalent to model predictive control, even for uncontrollable systems. Moreover, our results significantly reduce the amount of data needed in identifying homogeneous multi-agent systems.
\end{abstract}

\section{Introduction}
Willems' fundamental lemma provides a data-based parameterization of trajectories generated by linear time invariant (LTI) systems~\cite{willems2005note}. In particular, consider the LTI system
\begin{subequations}
\label{sys: LTI system}
\begin{align}
    x_{t+1}=&Ax_t+Bu_t,\label{eqn: dynamics AB}\\
    y_t=& Cx_t+Du_t,\label{eqn: dynamics CD}
\end{align}
\end{subequations}
where \(u_t\in\mathbb{R}^m,x_t\in\mathbb{R}^n, y_t\in\mathbb{R}^p\) denote the input, state and output of the system at discrete time \(t\), respectively. If system \eqref{sys: LTI system} is controllable, the lemma asserts that every length-\(L\) input-output trajectory of system \eqref{sys: LTI system} is a linear combination of a finite number of measured ones. These measured trajectories can be extracted from one single trajectory with persistent excitation of order \(n+L\) \cite{willems2005note}, or multiple trajectories with collective persistent excitation of order \(n+L\) \cite{van2020willems}. By parameterizing trajectories of the system \eqref{sys: LTI system} using measured data, the lemma has profound implications in system identification \cite{markovsky2005algorithms,katayama2006subspace,markovsky2019data}, and inspired a series of recent results including data-driven simulation \cite{markovsky2005algorithms,berberich2020trajectory}, output matching \cite{markovsky2008data}, control by interconnection \cite{maupong2017data}, set-invariance control \cite{bisoffi2019data}, linear quadratic regulation \cite{de2019formulas}, and predictive control \cite{huang2019data,coulson2019data,alpago2020extended,berberich2020data,allibhoy2020data,yin2020maximum,fabiani2020optimal}.

In the meantime, whether conditions of controllability and persistency of excitation in Willems' fundamental lemma are necessary has not been investigated in depth. The current work aims to address this issue by answering the following two questions. First, without controllability, to what extent can the linear combinations of a finite number of measured input-output trajectories parameterize all possible ones? Second, can the order of persistency of excitation be reduced? 

Recent results on system identification has shown that, assuming sufficient persistency of excitation, the linear combinations of a finite number of measured input-output trajectories contain any trajectory whose initial state is in the controllable subspace \cite{mishra2020data,markovsky2020identifiability}. As we show subsequently, such results only partially answer the first question above: trajectories with initial state outside the controllable subspace can also be contained in said linear combinations.

We answer the aforementioned questions by introducing extensions of Willems' fundamental lemma. First, we show that, with sufficient persistency of excitation, any length-\(L\) input-output trajectory whose initial state is in some subspace is a linear combination of a finite number of measured ones. Said subspace is the sum of the controllable subspace, unobservable subspace, and a certain subspace associated with the measured trajectories. Second, we show that the order of persistency of excitation required by Willems' fundamental lemma can be reduced from \(n+L\) to \(\delta_{\min}+L\), where \(\delta_{\min}\) is the degree of the minimal polynomial of the system matrix \(A\). Our first result completes those presented in \cite{mishra2020data,markovsky2020identifiability} by showing exactly which trajectories are parameterizable by a finite number of measured ones for an arbitrary LTI systems. Furthermore, this result shows that data-driven predictive control using online data is equivalent to model predictive control, not only for controllable systems, as shown in \cite{coulson2019data,alpago2020extended}, but also for uncontrollable ones. Our second result, compared with those in \cite{van2020willems}, can reduce the amount of data samples used in identifying homogeneous multi-agent systems by an order of magnitude.

The rest of the paper is organized as follows. We first prove an extended Willems' fundamental lemma and discuss its implications in Section~\ref{sec: theory}. We provide ramifications
of this extension for a representative set of applications in Section~\ref{sec: application} before providing concluding remarks in Section~\ref{sec: conclusion}.

\textit{Notation:} We let \(\mathbb
R\), and \(\mathbb{N}\) and \(\mathbb{N}_+\) denote the set of real numbers, non-negative integers, and positive integers, respectively. 
The image and right kernel of matrix \(M\) is denoted by \(\im M\) and \(\ker M\), respectively. Let \(\norm{x}_M=x^\top Mx\) for any matrix \(M\) and vector \(x\). 
When applied to subspaces, we let \(+\) and \(\times\) denote the sum \cite[p.2]{horn2012matrix} and Cartesian product operation \cite[p.370]{horn2012matrix}, respectively. We let \(\otimes\) denote the Kronecker product. Given a signal \(f:\mathbb{N}\to\mathbb{R}^{q}\) and \(i, j\in\mathbb{N}\) with \(i\leq j\), we denote 
\(f_{[i, j]}=\begin{bmatrix}
f_i^\top & f_{i+1}^\top & \cdots & f_{j}^\top
\end{bmatrix}^\top\), and the Hankel matrix of depth \(d\) (\(d \leq j-i+1 \)) associated with \(f_{[i, j]}\) as
\[
H_d(f_{[i, j]})=\begin{bmatrix}
f_i & f_{i+1} & \cdots & f_{j-d+1}\\
f_{i+1} & f_{i+2} & \cdots & f_{j-d+2}\\
\vdots & \vdots &  & \vdots \\
f_{i+d-1} & f_{i+d} & \cdots & f_{j}
\end{bmatrix}.
\]
\section{Extensions of Willems' fundamental lemma}
\label{sec: theory}
In this section, we introduce extensions of
Willems' fundamental lemma. Throughout we let
\begin{equation}\label{eqn: iso trajectory}
    (u^i_{[0, T^i-1]}, x^i_{[0, T^i-1]}, y^i_{[0, T^i-1]})
\end{equation}
denote a length-\(T^i\) (\(T^i\in\mathbb{N}_+\)) input-state-output trajectory generated by system \eqref{sys: LTI system} for all \(i=1, 2, \ldots, \tau\), where \(\tau\in\mathbb{N}_+\) is the total number of trajectories. We let
\begin{equation}\label{eqn: data trajectories}
\{u^i_{[0, T^i-1]}\}_{i=1}^\tau,\enskip \{x^i_{[0, T^i-1]}\}_{i=1}^\tau, \enskip  \{y^i_{[0, T^i-1]}\}_{i=1}^\tau,
\end{equation}
denote the set of input, state, and output trajectories, respectively. We will use the following subspaces,
\begin{equation}\label{eqn: subspace}
\begin{aligned}
    &\mathcal{R} = \im \begin{bmatrix}
        B & AB & \cdots & A^{n-1}B
    \end{bmatrix},\\
    &\mathcal{O} =  \ker \begin{bmatrix}
        C^\top & (CA)^\top & \cdots & (CA^{n-1})^\top
    \end{bmatrix}^\top,\\
    &\mathcal{K}[x_0^1, x_0^2, \ldots, x_0^\tau] = \im \begin{bmatrix}
    X_0 & AX_0 & \cdots & A^{n-1}X_0
    \end{bmatrix},
\end{aligned}
\end{equation}
where \(X_0=\begin{bmatrix} x_0^1 & x_0^2 & \cdots & x_0^\tau\end{bmatrix}\), and \(x_0^i\) is the first state in the trajectory \(x_{[0, T^i-1]}^i\) for all \(1\leq i\leq \tau\).
In particular, \(\mathcal{R}\) is known as the controllable subspace (we say that system \eqref{sys: LTI system} is controllable if \(\mathcal{R}=\mathbb{R}^n\)), \(\mathcal{O}\) is known as the unobservable subspace, and 
\(\mathcal{K}[x_0^1, x_0^2, \ldots, x_0^\tau]\) is the smallest \(A\)-invariant subspace containing \(x_0^1, x_0^2, \ldots, x_0^\tau\).  Using the Cayley-Hamilton theorem, one can verify that \eqref{sys: LTI system} ensures
\begin{equation}\label{eqn: reachable state}
    x_t=\textstyle A^t x_0+\sum_{j=0}^{t-1}A^{t-j-1}Bu_j \in \mathcal{R}+\mathcal{K}[x_0],
\end{equation}
for all \(t\in \mathbb{N}\), where \(\mathcal{K}[x_0]\) is defined similar to \(\mathcal{K}[x_0^1, x_0^2, \ldots, x_0^\tau]\). We let \(\delta\in\mathbb{N}_+\) be such that 
\begin{equation}\label{eqn: deg min poly}
    \delta\geq \delta_{\min}\coloneqq \text{degree of } p_{\min}(A),
\end{equation}
where \(p_{\min}(A)\) is the minimal polynomial of matrix \(A\) \cite[Def. 3.3.2]{horn2012matrix}, and, due to the Cayley-Hamilton theorem, \(\delta_{\min}\) is upper bounded by \(n\).
We will also use the following definitions
that streamline our subsequent analysis.

\begin{definition}
We say a length-\(L\) input-output trajectory \((\widebar{u}_{[0, L-1]}, \widebar{y}_{[0, L-1]})\), with \(L\in\mathbb{N}_+\),  is \emph{parameterizable} by \(\{u_{[0, T^i-1]}, y_{[0, T^i-1]}\}_{i=1}^\tau\) if there exists  \(g\in\mathbb{R}^{\sum_{i=1}^\tau(T^i-L+1)}\) such that
\begin{equation}
\label{eqn: Willems' II}
\begin{aligned}
    \begin{bmatrix}
\widebar{u}_{[0,L-1]}\\
\widebar{y}_{[0,L-1]}
\end{bmatrix}
= \begin{bmatrix}
      H_L(u^1_{[0, T^1-1]}) & \cdots & H_L(u^\tau_{[0, T^\tau-1]})\\
      H_L(y^1_{[0, T^1-1]}) & \cdots & H_L(y^\tau_{[0, T^\tau-1]})
      \end{bmatrix}g.
\end{aligned}
\end{equation}
\end{definition}

As an example, if \(\tau=2\), \(L=2\), \(T^1=3, T^2=4\), then  \eqref{eqn: Willems' II} assumes the form,
\begin{equation*}
\left[
    \begin{array}{c}
        \widebar{u}_0\\
\widebar{u}_1\\ \hdashline[2pt/2pt]
        \widebar{y}_0\\
\widebar{y}_1
    \end{array}
\right]=\left[
    \begin{array}{cc;{2pt/2pt}ccc}
       u_0^1 & u_1^1 & u_0^2 & u_1^2 & u_2^2\\
  u_1^1 & u_2^1 & u_1^2 & u_2^2 & u_3^2\\ \hdashline[2pt/2pt]
    y_0^1 & y_1^1 & y_0^2 & y_1^2 & y_2^2\\
  y_1^1 & y_2^1 & y_1^2 & y_2^2 & y_3^2
    \end{array}
\right]
g.
\end{equation*}

\begin{definition}[Collective persistent excitation \cite{van2020willems}]\label{def: persistent excitation}
We say that \(\{u_{[0, T^i-1]}\}_{i=1}^\tau\) is \emph{collectively persistently exciting} of order \(d\in\mathbb{N}_+\) if \(d\leq T^i\) for all \(i=1, 2, \ldots, \tau\), and the mosaic-Hankel matrix,
\begin{equation}\label{eqn: d mosaic-Hankel}
\begin{bmatrix}
    H_d(u^1_{[0, T^1-1]}) & \cdots & H_d(u^\tau_{[0, T^\tau-1]})
\end{bmatrix},
\end{equation}
has full row rank.
\end{definition}

If \(\tau=1\), then Definition~\ref{def: persistent excitation} reduces to the traditional notion of persistency of excitation \cite{willems2005note}. 

The Willems' fundamental lemma asserts that: if the system \eqref{sys: LTI system} is controllable and \(\{u^i_{[0, T^i-1]}\}_{i=1}^\tau\) is collectively persistently exciting of order \(n+L\), then \((\widebar{u}_{[0, L-1]}, \widebar{y}_{[0, L-1]})\) is parameterizable by \(\{u_{[0, T^i-1]}, y_{[0, T^i-1]}\}_{i=1}^\tau\) if and only if it is an input-output trajectory of the system \eqref{sys: LTI system} \cite{willems2005note,van2020willems}. As our main contribution, the following theorem shows that not only the controllability assumption can be relaxed, but also the required order of collective persistent excitation can be reduced from \(n+L\) to, at best, \(\delta_{\min}+L\).

\begin{theorem}\label{thm: Willem}
Let \(\{u^i_{[0, T^i-1]}, x^i_{[0, T^i-1]}, y^i_{[0, T^i-1]}\}_{i=1}^\tau\) be a set of input-state-output trajectories generated by the system \eqref{sys: LTI system}. If \(\{u_{[0, T^i-1]}\}_{i=1}^\tau\) is collectively persistently exciting of order \(\delta+L\) where \(\delta\) satisfies \eqref{eqn: deg min poly}, then
\begin{equation}\label{eqn: Willems' I}
    \begin{aligned}
     &\im\begin{bmatrix}
      H_1(x^1_{[0, T^1-L]}) & \cdots & H_1(x^\tau_{[0, T^\tau-L]})\\
      H_L(u^1_{[0, T^1-1]}) & \cdots & H_L(u^\tau_{[0, T^\tau-1]})
      \end{bmatrix}\\
      &= (\mathcal{R}+ \mathcal{K}[x^1_0, x^2_0, \ldots, x^\tau_0])\times \mathbb{R}^{mL}.
    \end{aligned}
\end{equation}
Further, \((\widebar{u}_{[0, L-1]}, \widebar{y}_{[0, L-1]})\) is parameterizable by \(\{u_{[0, T^i-1]}, y_{[0, T^i-1]}\}_{i=1}^\tau\) if and only if there exists a state trajectory \(\widebar{x}_{[0, L-1]}\) with 
\begin{equation}\label{eqn: set S}
    \widebar{x}_0\in \mathcal{R}+ \mathcal{O}+ \mathcal{K}[x_0^1, x_0^2, \ldots, x_0^\tau],
\end{equation}
such that \((\widebar{u}_{[0, L-1]}, \widebar{x}_{[0, L-1]}, \widebar{y}_{[0, L-1]})\) is an input-state-output trajectory generated by \eqref{sys: LTI system}. 
\end{theorem}

\begin{proof}
See the Appendix.
\end{proof}

\begin{remark}
The equality in \eqref{eqn: Willems' I} generalizes \cite[Lem. 2]{mishra2020data} by proving stronger results using weaker assumptions. Particularly, the assumption of \(n+L\) order of persistent excitation in \cite[Lem. 2]{mishra2020data} is reduced to \(\delta+L\) with \(\delta\geq \delta_{\min}\), and the controllable subspace \(\mathcal{R}\) used in \cite[Lem. 2]{mishra2020data} is extended to its superset \(\mathcal{R}+ \mathcal{K}[x_0^1, x_0^2, \ldots, x_0^\tau]\).
\end{remark}

Due to its dependence on state \(\widebar{x}_0\) and the unknown subspaces in \eqref{eqn: subspace}, the condition in \eqref{eqn: set S} is difficult to verify, especially if merely input-output trajectories are available. One approach to guarantee its satisfaction is to ensure the right hand side of \eqref{eqn: set S} equals \(\mathbb{R}^n\), either by assuming controllability (\ie, \(\mathcal{R}=\mathbb{R}^n\)) as in \cite{willems2005note} and \cite{van2020willems}, or requiring the input-state-output trajectories in \eqref{eqn: iso trajectory} satisfy \(\im \begin{bmatrix}x_0^1 & x_0^2& \ldots & x_0^\tau\end{bmatrix}=\mathbb{R}^n\). The following corollary, however, provides an alternative approach that requires neither controllability nor state measurements. 

\begin{corollary}\label{cor: online}
Let \((u_{[0, K-1]}, y_{[0, K-1]})\) be an input-output trajectory generated by the system \eqref{sys: LTI system}, and \(u_{[0, T-1]}\) be persistently exciting of order \(\delta+L\), where \(\delta\) satisfies \eqref{eqn: deg min poly} and \(L\leq T\leq K\).  Then \((u_{[t, t+L-1]}, y_{[t, t+L-1]})\) is parameterizable by \((u_{[0, T-1]}, y_{[0, T-1]})\) for all \(0\leq t\leq K-L\).
Conversely, if \((\widebar{u}_{[0, L-1]}, \widebar{y}_{[0, L-1]})\) is parameterizable by \((u_{[0, T-1]}, y_{[0, T-1]})\), then it is an input-output trajectory generated by the system \eqref{sys: LTI system}.
\end{corollary}

\begin{proof}
See the Appendix.
\end{proof}

Corollary~\ref{cor: online} shows that any length-\(L\) segment of an input-output trajectory is parameterizable by its first length-\(T\) segment, assuming sufficient persistency of excitation. 
This result is particularly useful in predictive control as we will show in Section~\ref{subsec: DPC}.

Another implication of Theorem~\ref{thm: Willem} is that the order of persistent excitation required by trajectory parameterization only depends on the degree of the minimal polynomial of matrix \(A\) in \eqref{sys: LTI system}, instead of its dimension. In general, it is difficult to establish a bound on the degree of the minimal polynomial of an unknown matrix tighter than its dimension. However, the following corollary shows an exception; its usefulness will be illustrated later in Section~\ref{subsec: multiagent}.

\begin{corollary}\label{cor: block diag}
If there exists \(\widebar{A}\in\mathbb{R}^{\widebar{n}\times \widebar{n}}\) such that \(A=I_N\otimes \widebar{A}\), where \(\widebar{n}, N\in\mathbb{N}_+\), then Theorem~\ref{thm: Willem} holds with \(\delta=\widebar{n}\).
\end{corollary}

The proof follows from Theorem~\ref{thm: Willem} and the fact that the minimal polynomial of \(A\) is the same as that of \(\widebar{A}\), which has degree at most \(\widebar{n}\).


\section{Applications}
\label{sec: application}
In this section, we provide two examples that illustrate the distinct implications of Theorem~\ref{thm: Willem}.
\subsection{Online data-enabled predictive control}
\label{subsec: DPC}


Model predictive control (MPC) provides an effective control strategy for systems with physical and operational constraints \cite{mayne2000constrained,mayne2014model}. In particular, consider system \eqref{sys: LTI system} with output feedback. At each sampling time \(t\in\mathbb{N}_+\), given the input-output measurements history \((u_{[0, t-1]}, y_{[0, t-1]})\), MPC solves the following optimization for input \(\widebar{u}_t\)
\begin{equation}\label{opt: MPC}
    \begin{array}{ll}
        \underset{\substack{\widebar{x}_{[t-N, t+L-1]}\\\widebar{u}_{[t, t+L-1]}\\\widebar{y}_{[t,t+L-1]} }}{\mbox{minimize}} & \sum_{k=t}^{t+L-1} \big(\norm{\widebar{y}_k-r_k}^2_Q+\norm{\widebar{u}_k}^2_R\big)\\
     \mbox{subject to} & \widebar{x}_{k+1}=A\widebar{x}_k+Bu_k, \enskip  y_k=C\widebar{x}_k+Du_k, \\
     & t-N\leq k\leq t-1, \\
     &\widebar{x}_{k+1}=A\widebar{x}_k+B\widebar{u}_k, \enskip  \widebar{y}_k=C\widebar{x}_k+D\widebar{u}_k, \\
     & t\leq k\leq t+L-1, \\
        & \widebar{y}_k\in\mathbb{Y}, \enskip \widebar{u}_k\in\mathbb{U},\enskip t\leq k\leq t+L-1,
    \end{array}
\end{equation}
where \(N\leq t\). The sets \(\mathbb{U}\subset\mathbb{R}^m\) and \(\mathbb{Y}\subset\mathbb{R}^p\) describe feasible inputs and outputs, respectively. Positive semi-definite matrices \(Q\in\mathbb{R}^{p\times p}\) and \(R\in\mathbb{R}^{m\times m}\), together with reference output trajectory \(r_{[t, t+L-1]}\in\mathbb{R}^{pL}\), define the output tracking cost function. The idea is to find a length-\((N+L)\) input-output trajectory that agrees with the past length-\(N\) measurements history \((u_{[t-N, t-1]}, y_{[t-N, t-1]})\) and optimizes the future length-\(L\) trajectory \((\widebar{u}_{[t, t+L-1]}, \widebar{y}_{[t, t+L-1]})\). If system \eqref{sys: LTI system} is observable and \(N\geq n\), the initial state of such a trajectory is uniquely determined \cite[Lem. 1]{markovsky2008data}.

Recently, \cite{coulson2019data,alpago2020extended} proposed a data-enabled predictive control (DeePC) that replaces optimization \eqref{opt: MPC} with 
\begin{equation}\label{opt: DPC}
    \begin{array}{ll}
        \underset{\substack{g, \widebar{u}_{[t, t+L-1]}\\
        \widebar{y}_{[t,t+L-1]}}}{\mbox{minimize}} & \sum_{k=t}^{t+L-1} \big(\norm{\widebar{y}_k-r_k}^2_Q+\norm{\widebar{u}_k}^2_R\big)\\
        \mbox{subject to} & \begin{bmatrix}
            u_{[t-N, t-1]}\\
            \widebar{u}_{[t, t+L-1]}\\
            y_{[t-N, t-1]}\\
            \widebar{y}_{[t, t+L-1]}
        \end{bmatrix}=\begin{bmatrix}
            H_{N+L}(u_{[0, T-1]})\\
            H_{N+L}(y_{[0, T-1]})
        \end{bmatrix}g,\\
        & \widebar{u}_k\in\mathbb{U},\enskip  \widebar{y}_k\in\mathbb{Y},\enskip t\leq k\leq t+L-1,
    \end{array}
\end{equation}
where \(T\leq t\). If system \eqref{eqn: dynamics AB} is controllable, Willems' fundamental lemma guarantees that optimization \eqref{opt: DPC} is equivalent to the one in \eqref{opt: MPC}, assuming \(u_{[0, T-1]}\) is sufficiently persistently exciting.

Unfortunately, such guarantee does not exist if the system \eqref{sys: LTI system} is uncontrollable, and is non-trivial to verify even if it is controllable. Particularly, verifying the controllability of the unknown system  \eqref{sys: LTI system} either requires \((A, B)\) having special zero patterns \cite{mayeda1979strong}, or computation using \((u_{[0, T-1]}, y_{[0, T-1]})\) that scale cubically with \(T\) \cite{mishra2020data}.

Fortunately, Corollary~\ref{cor: online} guarantees the equivalence between \eqref{opt: MPC} and \eqref{opt: DPC} by only assuming sufficient persistency of excitation on \(u_{[0, T-1]}\), regardless of the controllability of the system \eqref{sys: LTI system}. First, if \((\widebar{u}_{[t, t+L-1]}, \widebar{y}_{[t, t+L-1]})\) satisfies the constraints in \eqref{opt: MPC}, then
\[
\left(\begin{bmatrix}
u_{[t-N, t-1]}\\
\widebar{u}_{[t, t+L-1]}
\end{bmatrix}, \begin{bmatrix}
y_{[t-N, t-1]}\\
\widebar{y}_{[t, t+L-1]}
\end{bmatrix}\right) 
\]
is a segment of an input-output trajectory generated by the system \eqref{sys: LTI system} that starts with \((u_{[0, T-1]}, y_{[0, T-1]})\). Hence Corollary~\ref{cor: online} implies that \((\widebar{u}_{[t, t+L-1]}, \widebar{y}_{[t, t+L-1]})\) satisfies the constraints in \eqref{opt: DPC}. Second, Corollary~\ref{cor: online} also implies that any \((\widebar{u}_{[t, t+L-1]}, \widebar{y}_{[t, t+L-1]})\) satisfying the constraints in \eqref{opt: DPC} also satisfies those in \eqref{opt: MPC}. Therefore, the constraints of the optimizations in \eqref{opt: MPC} and \eqref{opt: DPC} are equivalent, and so are the optimizations themselves. Note that if we replace \((u_{[0, T-1]}, y_{[0, T-1]})\) in \eqref{opt: DPC} with an arbitrary input-output trajectory generated by the system \eqref{sys: LTI system} \emph{offline}, then the above equivalence may not hold. Hence we term optimization \eqref{opt: DPC} the \emph{online DeePC} problem.

\begin{figure}[!ht]
    \centering
    \includegraphics[width=0.9\linewidth]{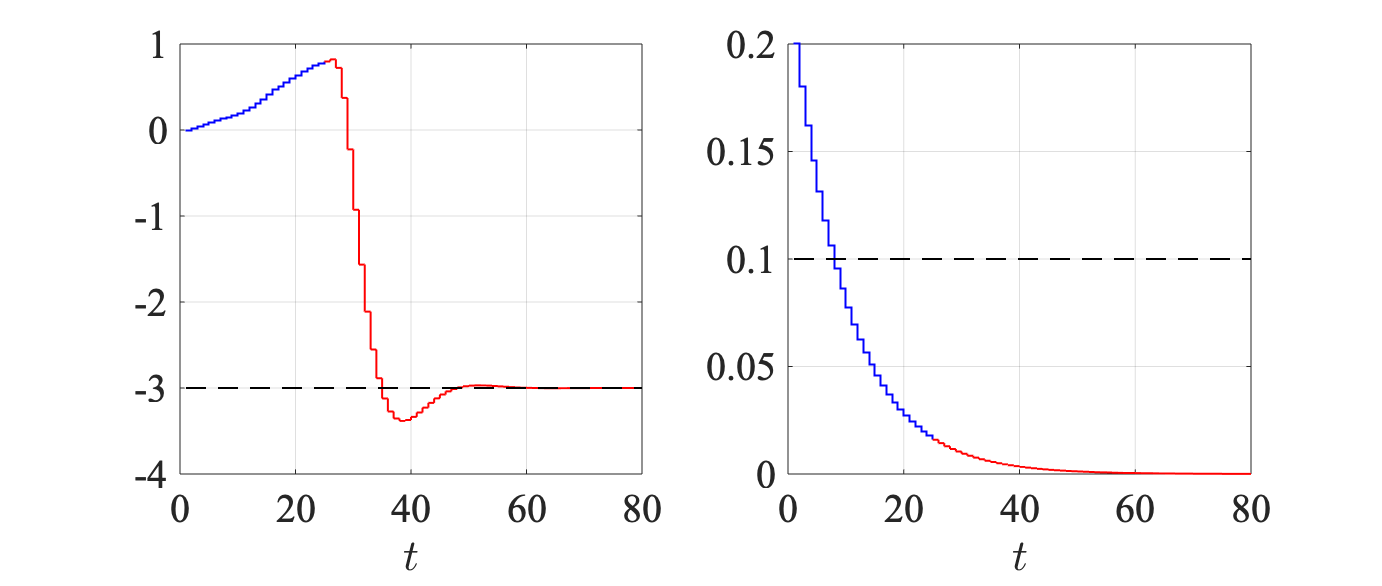}
    \caption{Output trajectory for Online DeePC. The blue and red curves denote \(y_{[0, T-1]}\) and \(y_{[T, K-1]}\), respectively. The dashed lines denote the output reference values.}
    \label{fig: online}
\end{figure}

To illustrate our results, we consider the system \eqref{sys: LTI system} where
\begin{equation}\label{sys: double integrator}
    A=\begin{bsmallmatrix}
    1 & 0.5 & 0 & 0\\
    0 & 1 & 0 & 0\\
    0 & 0 & 0.9 & 0.5\\
    0 & 0 & 0 & 0.9
    \end{bsmallmatrix},  B=\begin{bsmallmatrix}
    0.125\\
    0.5\\
    0\\
    0
    \end{bsmallmatrix}, C=\begin{bsmallmatrix}
    1 & 0\\
    0 & 0\\
    0 & 1\\
    0 & 0
    \end{bsmallmatrix}^\top, D=\begin{bsmallmatrix}
    0\\
    0
    \end{bsmallmatrix}.
\end{equation}
One can verify that the above \((A, B)\) is uncontrollable. We let \(N=4, L=5, T=25, K=80\) and generate online input-output trajectories \((u_{[0, T-1]}, y_{[0, T-1]})\), where \(u_t\) is sampled uniformly from \([-0.04, 0.04]\) for all \(t=0, \ldots, T-1\) such that \(u_{[0, T-1]}\) is persistently exciting of order \(n+L+N\). At time \(t=T, T+1, \ldots, K\), given \((u_{[0, t-1]}, y_{[0, t-1]})\), we obtain the input \(\widebar{u}_t\) by solving \eqref{opt: DPC}, where \(Q=I_2\), \(R=0.5\), \(\mathbb{Y}=\mathbb{R}^2\), \(\mathbb{U}=[-1, 1]\), and \(r_t=\begin{bmatrix} -3 & 0.1\end{bmatrix}^\top\) for all \(t\in\mathbb{N}\). Fig.~\ref{fig: online} shows that, although the system contains an uncontrollable output (right), online DeePC still ensures that the controllable output (left) tracks the reference value as desired. 

\subsection{Identification of homogeneous multi-agent systems}
\label{subsec: multiagent}
An important paradigm in formation control is the homogeneous multi-agent system, \ie, a network of \(N\) agents with the same LTI dynamics \cite{wen2014distributed,oh2015survey}. In such a system, each agent \(i\) can measure the state of agent \(j\) in a local coordinate system if \((i, j)\) is an edge of a directed graph \(\mathcal{G}\), which is composed of \(N\) nodes and \(M\) edges. The dynamics of this multi-agent system is given by \eqref{sys: LTI system} with
\begin{equation}\label{sys: multi-agent}
    A = I_{N}\otimes \widebar{A}, B=I_{N}\otimes \widebar{B}, C=E \otimes I_{\widebar{n}}, D=0_{M\widebar{n}\times N\widebar{m}},
\end{equation}
where \(\widebar{A}\in\mathbb{R}^{\widebar{n}\times \widebar{n}}\) and \(\widebar{B}\in\mathbb{R}^{\widebar{n}\times \widebar{m}}\) describe the input-state dynamics of an individual agent. Each row of matrix \(E\in\mathbb{R}^{M\times N}\) is indexed by a directed edge, \ie, an edge with a head and a tail, in graph \(\mathcal{G}\): the \(i\)-th entry in each row is ``\(1\)'' if node \(i\) is the head of the corresponding edge, ``\(-1\)'' if it is the tail, and ``\(0\)'' otherwise. We assume that \(\widebar{B}\) is a non-zero matrix and \((\widebar{A}, \widebar{B})\) is controllable. 

If at least one non-zero entry in matrix \(E\) is known, then system matrices in \eqref{sys: multi-agent} can be computed using the following \emph{Markov parameters} \cite[Sec. 3.4.4]{verhaegen2007filtering}
\begin{equation}\label{eqn: Markov param}
    \begin{aligned}
    M_k=& CA^{k-1}B+D\\
    =&(E \otimes I_{\widebar{n}})(I_{N}\otimes \widebar{A})^{k-1} (I_{N}\otimes \widebar{B})\\
    =& E \otimes (\widebar{A}^{k-1}\widebar{B}), \enskip \forall k=1,2, \ldots, \widebar{n}+1. 
    \end{aligned}
\end{equation}
In particular, let \((M_k)_{ij}\) denote the \(ij\)-th \(\widebar{n}\times \widebar{m}\) block of \(M_k\). If we know \(E_{ij}=1\) (the case of \(``-1"\) is similar), then \eqref{eqn: Markov param} implies \((M_k)_{ij}=\widebar{A}^{k-1}\widebar{B}\). For example, if \(E=\begin{bmatrix} 1 & -1
\end{bmatrix}\), then \eqref{eqn: Markov param} says \(M_k=\begin{bmatrix} \widebar{A}^{k-1}\widebar{B} & -\widebar{A}^{k-1}\widebar{B} \end{bmatrix}\). Hence given the Markov parameters \eqref{eqn: Markov param} and that \(E_{ij}=1\), we know \(E_{kl}\) is ``\(1\)'' if \((M_1)_{kl}=(M_1)_{ij}\), ``\(-1\)'' if \((M_1)_{kl}=-(M_1)_{ij}\), and ``0'' otherwise. Further, \(\widebar{B}=(M_1)_{ij}\) and  \(\widebar{A}\) is the unique solution to the following linear equations\footnote{Since \((\widebar{A}, \widebar{B})\) is controllable, matrix \(\begin{bmatrix}
   (M_1)_{ij}  & \cdots & (M_{\widebar{n}-1})_{ij}
    \end{bmatrix}=\begin{bmatrix}
   \widebar{B}  & \cdots & \widebar{A}^{\widebar{n}}\widebar{B}
    \end{bmatrix}\) has full column rank and \eqref{eqn: solve A bar} has a unique solution.}
\begin{equation}\label{eqn: solve A bar}
\begin{aligned}
    \widebar{A}\begin{bmatrix}
   (M_1)_{ij}  & \cdots & (M_{\widebar{n}})_{ij}
    \end{bmatrix}=\begin{bmatrix}
    (M_2)_{ij} & \cdots & (M_{\widebar{n}+1})_{ij}
    \end{bmatrix}.
\end{aligned}
\end{equation}

Therefore, given at least one non-zero entry in matrix \(E\), in order to compute the system matrices in \eqref{sys: multi-agent}, it suffices to know the Markov parameters \eqref{eqn: Markov param}. These parameters can be computed using Corollary~\ref{cor: block diag} via a data-driven simulation procedure \cite{markovsky2005algorithms,markovsky2008data} (see Appendix for details).

In numerical simulations, we consider the homogeneous multi-agent system used in \cite[Example 3]{wen2014distributed}, discretized with a sampling time of 0.1s such that the system dynamics is given by \eqref{sys: multi-agent} where
\begin{equation*}
    \widebar{A}=\begin{bsmallmatrix}
    0.9964 & 0.0026 & -0.0004 & -0.0460\\
    0.0045 & 0.9037 & -0.0188 & -0.3834\\
    0.0098 & 0.0339 & 0.9383 & 0.1302\\
    0.0005 & 0.0017 & 0.0968 & 1.0067
    \end{bsmallmatrix},\widebar{B}=\begin{bsmallmatrix}
    0.0445 & 0.0167\\
    0.3407 & -0.7249\\
    -0.5278 & 0.4214 \\
    -0.0268 & 0.0215
    \end{bsmallmatrix}.
\end{equation*} 
In addition, we let \(E=\begin{bmatrix}\mathbf{1}_{N-1} &
        -I_{N-1}
    \end{bmatrix}\), where \(\mathbf{1}_{N-1}\in\mathbb{R}^{N-1}\) is the vector of all \(1\)'s.

We compare Corollary~\ref{cor: block diag} against the results in \cite[Thm. 2]{van2020willems} in terms of the least amount of input-output data needed to compute matrices in \eqref{sys: multi-agent}. In particular, since matrix \(\widebar{A}\) has spectrum radius \(1.03\), we use use input-output trajectories \(\{u^i_{[0, T-1]}, y^i_{[0, T-1]}\}_{i=1}^\tau\) with relatively short length \(T=120\) to avoid numerical instability \cite{van2020willems}. The entries in \(\{u_{[0, T-1]}^i\}_{i=1}^\tau\) are sampled uniformly from \([-0.1, 0.1]\). Using Corollary~\ref{cor: block diag}, the data-driven simulation procedure requires \(\{u^i_{[0, T-1]}\}_{i=1}^\tau\) to be collectively persistently exciting of order \((N+1)\widebar{n}+1\). In other words, matrix \eqref{eqn: d mosaic-Hankel} with \(d=(N+1)\widebar{n}+1\)
has full row rank, hence it must have at least as many columns as rows, \ie,  \[\textstyle \tau\geq \frac{((N+1)\widebar{n}+1)N\widebar{m}}{T-(N+1)\widebar{n}}=\frac{8N^2+10N}{116-4N}.\]
In comparison, if we use \cite[Thm. 2]{van2020willems} instead of Corollary~\ref{cor: block diag}, we need \(\{u^i_{[0, T-1]}\}_{i=1}^\tau\) to be collectively persistently exciting of order \(2Nn+1\) (see \cite[Sec. IV-A]{van2020willems}).  In other words, matrix \eqref{eqn: d mosaic-Hankel} with \(d=2Nn+1\) has full row rank, which implies
\[\textstyle \tau\geq \frac{(2N\widebar{n}+1)N\widebar{m}}{T-2N\widebar{n}}=\frac{16N^2+2N}{120-8N}.\]

\begin{figure}[!ht]
    \centering
    \includegraphics[width=0.9\linewidth]{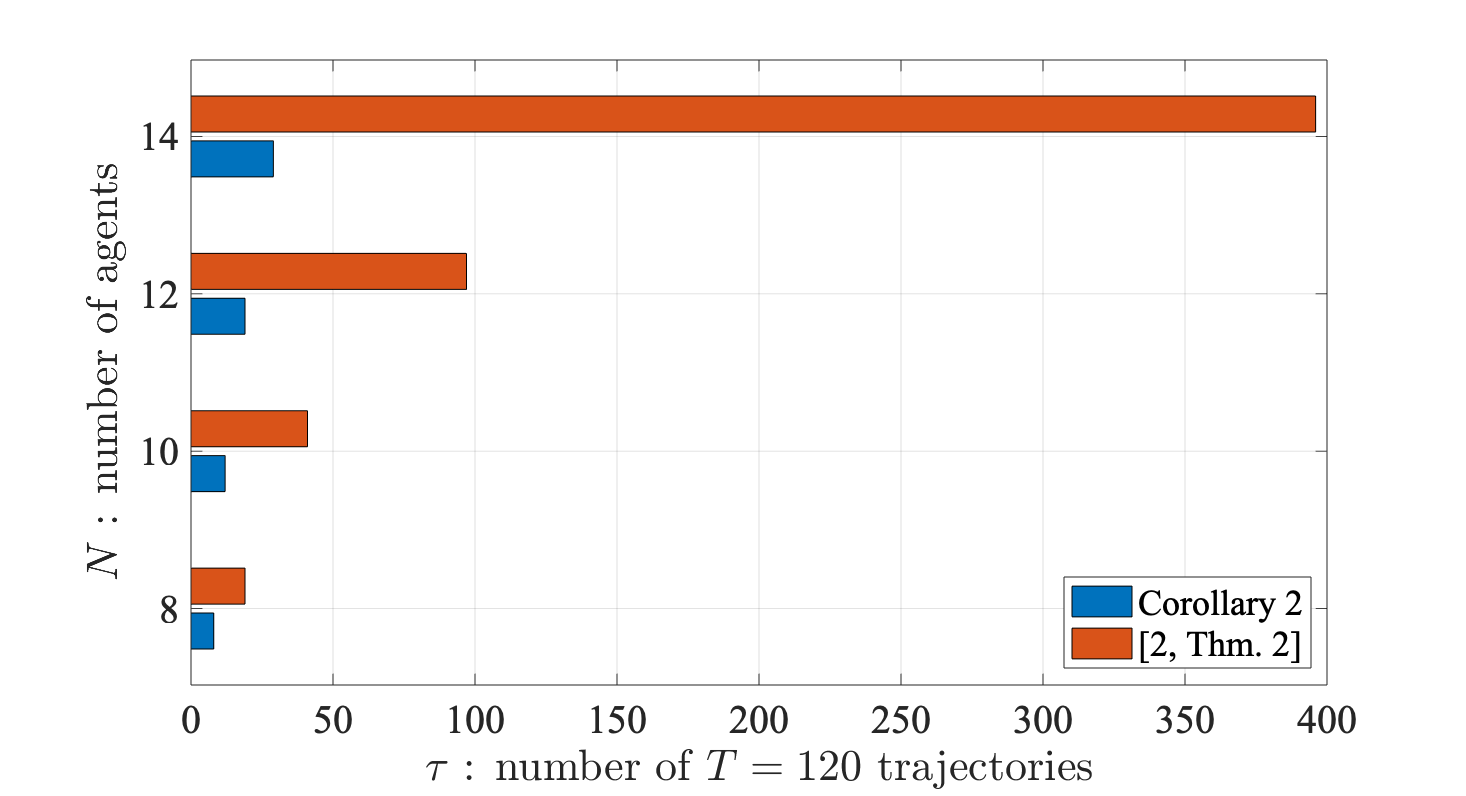}
    \caption{Minimum number of length \(T=120\) input-output trajectories required to identify system \eqref{sys: multi-agent}.}
    \label{fig: multiagent}
\end{figure}

In Fig.~\ref{fig: multiagent}, we show the minimum number of input-output trajectories required to compute matrices in \eqref{sys: multi-agent} in numerical simulations. The results tightly match the aforementioned two lower bounds, and the number of trajectories required by Corollary~\ref{cor: block diag} is one order of magnitude less than that of \cite[Thm. 1]{van2020willems} when \(N=14\).

\section{Conclusions}
\label{sec: conclusion}
We introduced an extension of Willems' fundamental lemma by relaxing the controllability and persistency of excitation assumptions. We demonstrate the usefulness of our results in the context of DeePC and identification of homogeneous multi-agent system. Future directions include generalizations to noisy data and nonlinear systems. 
\section*{Appendix}
\paragraph*{Proof of Theorem~\ref{thm: Willem}}
We start by proving the first statement using a double inclusion argument. Using an argument similar to \eqref{eqn: reachable state}, one can directly show that the left hand side of \eqref{eqn: Willems' I} is included in its right hand side. To show the other direction,
we show that the left kernel of matrix
\begin{equation}\label{eqn: Willems' I matrix}
    \begin{bmatrix}
      H_1(x^1_{[0, T^1-L]}) & \cdots & H_1(x^\tau_{[0, T^\tau-L]})\\
      H_L(u^1_{[0, T^1-1]}) & \cdots & H_L(u^\tau_{[0, T^\tau-L]})
      \end{bmatrix}
\end{equation}
is orthogonal to \((\mathcal{R}+ \mathcal{K}[x_0^1, x_0^2, \ldots, x_0^\tau]) \times \mathbb{R}^{mL}\).\textbf{}
To this end, let 
\begin{equation}\label{eqn: kernel element}
    v^\top=\begin{bmatrix} \xi^\top & \eta_1^\top & \eta_2^\top & \cdots \eta_L^\top \end{bmatrix},
\end{equation}
be an arbitrary row vector in the left kernel of matrix \eqref{eqn: Willems' I matrix}, where \(\xi\in\mathbb{R}^n, \eta_1, \eta_2, \ldots, \eta_L \in\mathbb{R}^{m}\). Since \(\delta\geq \delta_{\min}\), using \cite[Def. 3.3.2]{horn2012matrix}, we know there exists \(\alpha_{0k}, \alpha_{1k}, \ldots,\alpha_{\delta-1, k}\in\mathbb{R}\) such that
\begin{equation}\label{eqn: min poly}
    A^k+\textstyle \sum_{j=0}^{\delta-1} \alpha_{jk}A^j=0_{n\times n}, \enskip \forall k= \delta, \delta+1, \ldots 
\end{equation}
The above equation implies that \(A^kB=\textstyle -\sum_{j=0}^{\delta-1}\alpha_{jk}A^jB\) and \(A^kx_0^i=\textstyle -\sum_{j=0}^{\delta-1}\alpha_{jk}A^jx_0^i\) for all \(k=\delta, \ldots, n-1\) and \( i=1, \ldots, \tau.\)
Therefore in order to show the left kernel of matrix \eqref{eqn: Willems' I matrix} is orthogonal to \((\mathcal{R}+ \mathcal{K}[x_0^1, x_0^2, \ldots, x_0^\tau])\times \mathbb{R}^{mL}\), it suffices to show the following
\begin{subequations}
\begin{align}
    \eta_1^\top=&\eta_2^\top=\cdots=\eta_L^\top=0_m^\top, \label{eqn: kernel 1}\\
    \xi^\top B=&\xi^\top AB=\cdots=\xi^\top A^{\delta-1}B=0_m^\top,\label{eqn: kernel 2}\\
    \xi^\top x^i_0=&\xi^\top Ax^i_0=\cdots =\xi^\top A^{\delta-1} x^i_0=0.\enskip \forall i=1, \ldots,\tau.\label{eqn: kernel 3}
\end{align}
\end{subequations}

In order to prove \eqref{eqn: kernel 1}, we let \(w_0, w_1, \ldots, w_{\delta}\in\mathbb{R}^{n+m(\delta+L)}\) be such that \(w_0=\begin{bmatrix}v^\top & 0_{m\delta}^\top\end{bmatrix}^\top\) and \(w_j\) equals
\begin{gather*}
    \begin{bmatrix}
    \xi^\top A^j & \xi^\top A^{j-1}B & \cdots \xi^\top B & \eta_1^\top & \cdots & \eta_L^\top & 0_{m(\delta-j)}^\top 
    \end{bmatrix}^\top,
\end{gather*}
for \(j=1, \ldots, \delta\). Since \(v^\top \) is in the left kernel of matrix \eqref{eqn: Willems' I matrix}, using \eqref{sys: LTI system} one can verify that \(w_0^\top, w_1^\top, \ldots, w_{\delta}^\top \) are in the left kernel of
\begin{equation}\label{eqn: deeper Hankel}
    \begin{bmatrix}
    H_1(x^1_{[0, T^1-\delta-L]}) & \cdots & H_1(x^\tau_{[0, T^\tau-\delta-L]}) \\
    H_{\delta+L}(u^1_{[0, T^1-1]}) & \cdots & H_{\delta+L}(u^\tau_{[0, T^\tau-1]})
    \end{bmatrix}.
\end{equation}
Let \(\alpha_{\delta\delta}=1\) and \(k=\delta\) in \eqref{eqn: min poly}, 
we have \(0_{n\times n}=\sum_{j=0}^{\delta} \alpha_{j\delta}A^j\).
Hence
\begin{equation}\label{eqn: alpha w}
    \textstyle\sum_{j=0}^{\delta}\alpha_{j\delta} w_j^\top=\begin{bmatrix}
    \sum_{j=0}^{\delta} \alpha_{j\delta}\xi^\top A^j & r^\top
    \end{bmatrix}=\begin{bmatrix}
    0_n & r^\top
    \end{bmatrix},
\end{equation}
for some vector \(r\in\mathbb{R}^{m(\delta+L)}\). Since row vectors \(w_0^\top, w_1^\top, \ldots, w_{\delta}^\top \) are in the left kernel of matrix \eqref{eqn: deeper Hankel}, equation \eqref{eqn: alpha w} implies that \(r^\top\) is in the left kernel of matrix 
\begin{equation}\label{eqn: mosaic-Hankel}
\begin{bmatrix}
    H_{\delta+L}(u^1_{[0, T^i-1]}) & \cdots & H_{\delta+L}(u^\tau_{[0, T^i-1]})
\end{bmatrix}.
\end{equation}
Since \(\{u_{[0, T^i-1]}^i\}_{i=1}^\tau\) is collectively persistently exciting of order \(\delta+L\), matrix \eqref{eqn: mosaic-Hankel} has full row rank. Therefore
\begin{equation}\label{eqn: trivial kernel}
    r=0_{m(\delta+L)}.
\end{equation}
Observe that the last \(m\) entries of \(r\) are given by
\(\alpha_{\delta\delta}\eta_L=\eta_L\), hence equation \eqref{eqn: trivial kernel} implies that \(\eta_L=0_m\). Then the last \(2m\) entries of \(r\) are given by \(\begin{bmatrix}
\alpha_{\delta\delta}\eta_{L-1}^\top+\alpha_{(\delta-1)\delta}\eta_{L}^\top & \alpha_{\delta\delta}\eta_{L}^\top
\end{bmatrix}^\top \). Since \(\eta_L=0_{m}\) and \(\alpha_{\delta\delta}=1\), equation \eqref{eqn: trivial kernel} also implies that \(\eta_{L-1}=0_m\). By repeating similar induction we can prove that \eqref{eqn: kernel 1} holds.

Next, since \eqref{eqn: kernel 1} holds, the first \(m\delta\) entries in \(r\) are
\begin{gather*}
    \begin{bmatrix}
     \sum\limits_{j=1}^{\delta}\alpha_{j\delta}\xi^\top A^{j-1}B & \sum\limits_{j=2}^{\delta}\alpha_{j\delta}\xi^\top A^{j-2}B & \cdots & \alpha_{\delta\delta}\xi^\top B
    \end{bmatrix}^\top .
\end{gather*}
By combining this with \eqref{eqn: trivial kernel} we get that
\begin{equation}\label{eqn: xi AB}
   \textstyle  0_m^\top=\sum_{j=k}^{\delta} \alpha_{j\delta}\xi^\top A^{j-k}B,\enskip \forall k=1, \ldots, \delta.
\end{equation}
Since \(\alpha_{\delta\delta}=1\), considering \(k=\delta\) in \eqref{eqn: xi AB} implies that \(\xi^\top B=0_m\). Substitute this back into \eqref{eqn: xi AB} and considering \(k=\delta-1\) implies that \(\xi^\top AB=0_m\). By repeating a similar reasoning for \(k=\delta-2, \delta-3, \ldots, 1\) we can prove that \eqref{eqn: kernel 2} holds. 

Further, by using \eqref{eqn: min poly} and \eqref{eqn: kernel 2} we can show that \(\xi^\top A^kB=0_m\) for all \( k\geq 1\). Combining this together with the fact that row vector \eqref{eqn: kernel element} is in the left kernel of matrix \eqref{eqn: Willems' I matrix} and that \eqref{eqn: kernel 1} also holds, it follows that,
\begin{equation*}
\begin{aligned}
    0=&\xi^\top x_k^i=\textstyle \xi^\top \big(A^kx^i_0+\sum_{j=0}^{k-1}A^{k-j-1}Bu^i_j\big)= \xi^\top A^kx^i_0,\\
    &\forall k=0, \ldots, T^i-L, \enskip i=1, \ldots, \tau.
\end{aligned}
\end{equation*}
Since \(T^i\geq \delta+L\) for \(i=1, \ldots, \tau\) by assumption, we conclude that \eqref{eqn: kernel 3} holds.

We now prove the second statement. Given the input, state, and output trajectories in \eqref{eqn: data trajectories}, suppose \eqref{eqn: Willems' II} holds. Let
\begin{equation}\label{eqn: state traj}
    \begin{aligned}
        \widebar{x}_0=&\begin{bmatrix}
H_1(x^1_{[0, T^1-L]}) & \cdots & H_1(x^\tau_{[0, T^\tau-L]})
\end{bmatrix}g,\\
\widebar{x}_{t+1}=&A\widebar{x}_t+B\widebar{u}_t, \enskip 0\leq t\leq L-2.
    \end{aligned}
\end{equation}
Then from \eqref{eqn: Willems' I} we know \(\widebar{x}_0\) satisfies \eqref{eqn: set S}, and one can verify that \((\widebar{u}_{[0, L-1]}, \widebar{x}_{[0, L-1]}, \widebar{y}_{[0, L-1]})\) is indeed an input-state-output trajectory of system \eqref{sys: LTI system}.

Conversely, let \((\widebar{u}_{[0, L-1]}, \widebar{x}_{[0, L-1]}, \widebar{y}_{[0, L-1]})\) be an input-state-output trajectory of system \eqref{sys: LTI system} with \(\widebar{x}_0\in\mathcal{R}+ \mathcal{O}+ \mathcal{K}[x_0^1, x_0^2, \ldots, x_0^\tau]\). Then there exists
\begin{equation}\label{eqn: state decomp}
    \widebar{x}_0^a\in\mathcal{R}+ \mathcal{K}[x_0^1, x_0^2, \ldots, x_0^\tau], \enskip \widebar{x}_0^b\in\mathcal{O},
\end{equation}
such that \(\widebar{x}_0=\widebar{x}_0^a+\widebar{x}_0^b\) and
\begin{equation}\label{eqn: thm1 eqn1}
    \begin{bmatrix}
\widebar{u}_{[0, L-1]}\\
\widebar{y}_{[0, L-1]}
\end{bmatrix}=\begin{bmatrix}
0  & I\\
O_L & T_L
\end{bmatrix}\begin{bmatrix}
\widebar{x}_0^a+\widebar{x}_0^b\\
\widebar{u}_{[0, L-1]}
\end{bmatrix}
\end{equation}
where 
\begin{equation}\label{eqn: T & O matrix}
\begin{aligned}
 T_L=&\begin{bmatrix}
    D & 0  & 0 & \cdots & 0 \\
    CB & D  & 0 & \cdots & 0\\
    CAB & CB  & D & \cdots & 0 \\
    \vdots &  \vdots & \vdots & \ddots & \vdots\\
    CA^{L-2}B & CA^{L-3}B & CA^{L-4}B & \cdots & D 
    \end{bmatrix}, \\
    O_L=&\begin{bmatrix}
    C^\top &
    (CA)^\top&
    (CA^2)^\top&
    \cdots &
    (CA^{L-1})^\top
    \end{bmatrix}^\top.
\end{aligned}
\end{equation}
Further, using the Cayley-Hamilton theorem one can show that \(\mathcal{O}\subset \ker O_L\) for any \(L\in\mathbb{N}_+\). Hence \eqref{eqn: state decomp} implies 
\begin{equation}\label{eqn: thm1 eqn2}
    O_L x_0^b=0_{Lp}.
\end{equation}
Since \(\widebar{x}_0^a\in \mathcal{R}+ \mathcal{K}[x_0^1, x_0^2, \ldots, x_0^\tau]\), the first statement of Theorem~\ref{thm: Willem} implies that there exists \(g\in\mathbb{R}^{\sum_{i=1}^\tau(T^i-L+1)}\) such that
\begin{equation}\label{eqn: thm1 eqn3}
    \begin{bmatrix}
\widebar{x}_0^a\\
\widebar{u}_{[0, L-1]}
\end{bmatrix}=\begin{bmatrix}
      H_1(x^1_{[0, T^1-L]}) & \cdots & H_1(x^\tau_{[0, T^\tau-L]})\\
      H_L(u^1_{[0, T^1-1]}) & \cdots & H_L(u^\tau_{[0, T^\tau-1]})
      \end{bmatrix}g.
\end{equation}
Notice that
\begin{equation}\label{eqn: thm1 eqn4}
\begin{aligned}
&\begin{bmatrix}
0  & I\\
O_L & T_L
\end{bmatrix}\begin{bmatrix}
      H_1(x^i_{[0, T^i-L]})\\
      H_L(u^i_{[0, T^i-1]})
      \end{bmatrix}=\begin{bmatrix}
        H_L(u^i_{[0, T^i-1]})  \\
            H_L(y^i_{[0, T^i-1]}) 
        \end{bmatrix}.
\end{aligned}        
\end{equation}
for \(i=1, \ldots, \tau\).
Substituting \eqref{eqn: thm1 eqn2}, \eqref{eqn: thm1 eqn3} and \eqref{eqn: thm1 eqn4} into \eqref{eqn: thm1 eqn1} gives \eqref{eqn: Willems' II}, 
thus completing the proof. \qed

\paragraph*{Proof of Corollary~\ref{cor: online}}
First, let \(x_{[0, K-1]}\) be such that \((u_{[0, K-1]}, x_{[0, K-1]}, y_{[0, K-1]})\) is an input-state-output trajectory generated by the system \eqref{sys: LTI system}. Then \eqref{eqn: reachable state} holds for all \(0\leq t\leq K-1\). Hence \((u_{[t, t+L-1]}, x_{[t, t+L-1]}, y_{[t, t+L-1]})\) is an input-state-output trajectory generated by system \eqref{sys: LTI system} that satisfies \eqref{eqn: reachable state}. From the second statement in Theorem~\ref{thm: Willem} we know that \((u_{[t, t+L-1]}, y_{[t, t+L-1]})\) is parameterizable by \((u_{[0, T-1]}, y_{[0, T-1]})\). 
Second, if \((\widebar{u}_{[0, L-1]}, \widebar{y}_{[0, L-1]})\) is parameterizable by \((u_{[0, T-1]}, y_{[0, T-1]})\), then one can verify that it is indeed an input-output trajectory generated by the system  \eqref{sys: LTI system} by constructing a length-\(L\) state trajectory similar to the one in \eqref{eqn: state traj}. 
\qed

\paragraph*{Markov parameters computation in Section~\ref{subsec: multiagent}}
Let \(\{u^i_{[0, T^i-1]}, y^i_{[0, T^i-1]}\}_{i=1}^\tau\) be input-output trajectories of the system described by \eqref{sys: multi-agent}, such that inputs \(\{u^i_{[0, T^i-1]}\}_{i=1}^\tau\) are collectively persistently exciting of order \((N+1)\widebar{n}+1\). Let \(n=N\widebar{n}, m=N\widebar{m}, p=M\widebar{n}\). Since \((\widebar{A},\widebar{B})\) in \eqref{sys: multi-agent} is controllable, one can verify that \(\mathcal{R}+ \mathcal{K}[x_0^1, x_0^2, \ldots, x_0^\tau]=\mathbb{R}^{n}\), regardless of the values of \(x^1_0, x_0^2, \ldots, x^\tau_0\). Using Corollary~\ref{cor: block diag} we know that there exists a matrix \(G_k\in\mathbb{R}^{(\sum_{i=1}^\tau (T^i-n))\times m}\) such that,
\begin{equation}\label{eqn: impulse}
\begin{aligned}
    &\begin{bmatrix}
    H_{n+1}(u^1_{[0, T^1-1]}) & \cdots & H_{n+1}(u^\tau_{[0, T^\tau-1]})\\
    H_{n+1}(y^1_{[0, T^1-1]}) & \cdots & H_{n+1}(y^\tau_{[0, T^\tau-1]})
    \end{bmatrix}G_k\\
    =&
    \begin{bmatrix}
0^\top_{m(n-k)\times m} & I_m & 0^\top_{(pn+km-kp)\times m} & M_0^\top & \cdots & M_k^\top
\end{bmatrix}^\top
\end{aligned}    
\end{equation}
for all \(k=0, 1,\ldots, \widebar{n}+1\), where \(M_0=0_{p\times m}\) and \(M_k\) is given by \eqref{eqn: Markov param}; also see \cite[Sec. 4.5]{markovsky2008data} for details. Next, given \(M_0, \ldots, M_{k-1}\), we can compute \(M_k\) by first solving the first \(m(n+1)+pn\) equations in \eqref{eqn: impulse} for matrix \(G_k\), then substituting the solution into the last \(p\) equations in \eqref{eqn: impulse}. By repeating this process for \(k=1, \ldots, \widebar{n}+1\) we obtain Markov parameters \eqref{eqn: Markov param}. Using Kalman decomposition one can verify that Markov parameters of system \eqref{sys: multi-agent} are the same as those of a reduced order controllable and observable system with state dimension less than \(n\). Hence matrix \(M_k\) obtained this way is unique; see \cite[Prop. 1]{markovsky2008data}.




\bibliographystyle{IEEEtran}
\bibliography{IEEEabrv,reference}

\begin{thebibliography}{10}
\providecommand{\url}[1]{#1}
\csname url@rmstyle\endcsname
\providecommand{\newblock}{\relax}
\providecommand{\bibinfo}[2]{#2}
\providecommand\BIBentrySTDinterwordspacing{\spaceskip=0pt\relax}
\providecommand\BIBentryALTinterwordstretchfactor{4}
\providecommand\BIBentryALTinterwordspacing{\spaceskip=\fontdimen2\font plus
\BIBentryALTinterwordstretchfactor\fontdimen3\font minus
  \fontdimen4\font\relax}
\providecommand\BIBforeignlanguage[2]{{%
\expandafter\ifx\csname l@#1\endcsname\relax
\typeout{** WARNING: IEEEtran.bst: No hyphenation pattern has been}%
\typeout{** loaded for the language `#1'. Using the pattern for}%
\typeout{** the default language instead.}%
\else
\language=\csname l@#1\endcsname
\fi
#2}}

\bibitem{willems2005note}
J.~C. Willems, P.~Rapisarda, I.~Markovsky, and B.~L. De~Moor, ``A note on
  persistency of excitation,'' \emph{Syst. Control Lett.}, vol.~54, no.~4, pp.
  325--329, 2005.

\bibitem{van2020willems}
H.~J. van Waarde, C.~De~Persis, M.~K. Camlibel, and P.~Tesi, ``Willems’
  fundamental lemma for state-space systems and its extension to multiple
  datasets,'' \emph{IEEE Control Syst. Lett.}, vol.~4, no.~3, pp. 602--607,
  2020.

\bibitem{markovsky2005algorithms}
I.~Markovsky, J.~C. Willems, P.~Rapisarda, and B.~L. De~Moor, ``Algorithms for
  deterministic balanced subspace identification,'' \emph{Automatica}, vol.~41,
  no.~5, pp. 755--766, 2005.

\bibitem{katayama2006subspace}
T.~Katayama, \emph{Subspace methods for system identification}.\hskip 1em plus
  0.5em minus 0.4em\relax Springer Science \& Business Media, 2006.

\bibitem{markovsky2019data}
I.~Markovsky, ``From data to models,'' in \emph{Low-Rank Approximation}.\hskip
  1em plus 0.5em minus 0.4em\relax Springer, 2019, pp. 37--70.

\bibitem{berberich2020trajectory}
J.~Berberich and F.~Allg{\"o}wer, ``A trajectory-based framework for
  data-driven system analysis and control,'' in \emph{Eur. Control Conf.}\hskip
  1em plus 0.5em minus 0.4em\relax IEEE, 2020, pp. 1365--1370.

\bibitem{markovsky2008data}
I.~Markovsky and P.~Rapisarda, ``Data-driven simulation and control,''
  \emph{Int. J. Control}, vol.~81, no.~12, pp. 1946--1959, 2008.

\bibitem{maupong2017data}
T.~Maupong and P.~Rapisarda, ``Data-driven control: A behavioral approach,''
  \emph{Syst. Control Lett.}, vol. 101, pp. 37--43, 2017.

\bibitem{bisoffi2019data}
A.~Bisoffi, C.~De~Persis, and P.~Tesi, ``Data-based guarantees of set
  invariance properties,'' \emph{arXiv preprint arXiv:1911.12293[eecs.SY]},
  2019.

\bibitem{de2019formulas}
C.~De~Persis and P.~Tesi, ``Formulas for data-driven control: Stabilization,
  optimality, and robustness,'' \emph{IEEE Trans. Autom. Control}, vol.~65,
  no.~3, pp. 909--924, 2019.

\bibitem{huang2019data}
L.~Huang, J.~Coulson, J.~Lygeros, and F.~D{\"o}rfler, ``Data-enabled predictive
  control for grid-connected power converters,'' in \emph{IEEE Conf. Decision
  Control}.\hskip 1em plus 0.5em minus 0.4em\relax IEEE, 2019, pp. 8130--8135.

\bibitem{coulson2019data}
J.~Coulson, J.~Lygeros, and F.~D{\"o}rfler, ``Data-enabled predictive control:
  In the shallows of the deepc,'' in \emph{Eur. Control Conf.}\hskip 1em plus
  0.5em minus 0.4em\relax IEEE, 2019, pp. 307--312.

\bibitem{alpago2020extended}
D.~{Alpago}, F.~{Dörfler}, and J.~{Lygeros}, ``An extended {K}alman filter for
  data-enabled predictive control,'' \emph{IEEE Control Systems Letters},
  vol.~4, no.~4, pp. 994--999, 2020.

\bibitem{berberich2020data}
J.~{Berberich}, J.~{Koehler}, M.~A. {Muller}, and F.~{Allgower}, ``Data-driven
  model predictive control with stability and robustness guarantees,''
  \emph{IEEE Transactions on Automatic Control}, pp. 1--1, 2020.

\bibitem{allibhoy2020data}
A.~Allibhoy and J.~Cort{\'e}s, ``Data-based receding horizon control of linear
  network systems,'' \emph{arXiv preprint arXiv:2003.09813}, 2020.

\bibitem{yin2020maximum}
M.~Yin, A.~Iannelli, and R.~S. Smith, ``Maximum likelihood estimation in
  data-driven modeling and control,'' \emph{arXiv preprint arXiv:2011.00925
  [eess.SY]}, 2020.

\bibitem{fabiani2020optimal}
F.~Fabiani and P.~J. Goulart, ``The optimal transport paradigm enables data
  compression in data-driven robust control,'' \emph{arXiv preprint
  arXiv:2005.09393 [eess.SY]}, 2020.

\bibitem{mishra2020data}
V.~K. Mishra, I.~Markovsky, and B.~Grossmann, ``Data-driven tests for
  controllability,'' \emph{IEEE Control Syst. Lett.}, vol.~5, no.~2, pp.
  517--522, 2020.

\bibitem{markovsky2020identifiability}
I.~Markovsky and F.~D\"{o}rfler, ``Identifiability in the behavioral setting,''
  Dept. ELEC, Vrije Universiteit Brussel, Tech. Rep., 2020.

\bibitem{horn2012matrix}
R.~A. Horn and C.~R. Johnson, \emph{Matrix analysis}.\hskip 1em plus 0.5em
  minus 0.4em\relax Cambridge university press, 2012.

\bibitem{mayne2000constrained}
D.~Q. Mayne, J.~B. Rawlings, C.~V. Rao, and P.~O. Scokaert, ``Constrained model
  predictive control: Stability and optimality,'' \emph{Automatica}, vol.~36,
  no.~6, pp. 789--814, 2000.

\bibitem{mayne2014model}
D.~Q. Mayne, ``Model predictive control: Recent developments and future
  promise,'' \emph{Automatica}, vol.~50, no.~12, pp. 2967--2986, 2014.

\bibitem{mayeda1979strong}
H.~Mayeda and T.~Yamada, ``Strong structural controllability,'' \emph{SIAM
  Journal on Control and Optimization}, vol.~17, no.~1, pp. 123--138, 1979.

\bibitem{wen2014distributed}
G.~Wen, Z.~Duan, W.~Ren, and G.~Chen, ``Distributed consensus of multi-agent
  systems with general linear node dynamics and intermittent communications,''
  \emph{Int. J. Robust. Nonlinear Control}, vol.~24, no.~16, pp. 2438--2457,
  2014.

\bibitem{oh2015survey}
K.-K. Oh, M.-C. Park, and H.-S. Ahn, ``A survey of multi-agent formation
  control,'' \emph{Automatica}, vol.~53, pp. 424--440, 2015.

\bibitem{verhaegen2007filtering}
M.~Verhaegen and V.~Verdult, \emph{Filtering and system identification: a least
  squares approach}.\hskip 1em plus 0.5em minus 0.4em\relax Cambridge
  University Press, 2007.

\end{thebibliography}

\end{document}